\documentclass[11pt]{article} 
\usepackage{amsfonts,amscd,amsmath}

\def\NP#1#2{ Nucl.Phys. B#1 (#2)} 
\def\PL#1#2{ Phys.Lett. B#1 (#2)}

\def\PR#1#2{Phys.Rev. D#1 (#2)} 
\def\IJMP#1#2{ Int.J.Mod.Phys. A#1 (#2)}

\def\HP#1#2{ JHEP #1 (#2)} 
\def\pd{\partial}

\def\nv{\frac{1}{1+f}\,t}
\def\tr{\text{tr}}
\newcommand{\ep}{\text e}
\newcommand{\oh}{\frac{1}{2}}
\title{Some Computations of Partition Functions and Tachyon Potentials in Background 
Independent Off-Shell String Theory}
\author{Oleg Andreev\thanks{e-mail:  andreev@physik.hu-berlin.de}
\thanks{Permanent address: Landau Institute for Theoretical Physics, Moscow, Russia}
\\ \\
Humboldt--Universit\"at zu Berlin, Institut f\"ur Physik\\
Invalidenstra\ss e 110, D-10115 Berlin, Germany}

\date{}
\begin{document} 
 
\maketitle 
\begin{abstract} 
We discuss what information can be safely extracted from background independent off-shell string 
theory. The major obstacle in doing so is that renormalization conditions of the underlying 
world-sheet theories are not exactly known. To get some insight, we first consider the tachyon and gauge 
field backgrounds and carry out computations in different renormalization schemes for both, bosonic 
string and superstring. Next, we use a principle of universality (renormalization scheme independence) to 
somehow compensate the missing of the renormalization conditions and get information we are 
looking for. It turns out that some asymptotics which are responsible for the potentials only obey 
the principle of universality.   
\end{abstract}

\vspace{-13cm}
\begin{flushright}
hep-th/0010218       \\
HU Berlin-EP-00/43
\end{flushright}
\vspace{11cm}

%\vspace{-5cm}
%_______________________     I N T R O D U C T I O N_________________
%\vspace{-.75cm}
\section{ Introduction} 
\renewcommand{\theequation}{1.\arabic{equation}}
\setcounter{equation}{0}

The background independent open string theory (BIOST)\cite{W,LW,S,S1} turns out to be a powerful tool to 
compute the exact tree level tachyon potentials \cite{GS,KMM,susy} which amusingly coincide with 
the potentials found within toy models based on the exactly solvable Schroedinger problem \cite{MZ,susyZ}. 
Such a theory essentially uses two-dimensional world-sheet theories on the disk which are free field theories 
inside the disk but includes non-trivial interactions at the boundary \footnote{For earlier discussions 
of world-sheet theories see, e.g., \cite{sigma} and references therein.}. 

From the beginning, it was emphasized in \cite{W} that the construction of BIOST is rather formal as it 
ignores ultraviolet divergences associated with unrenormalizable world-sheet interactions. Moreover, 
it is not known precisely what the renormalization conditions for world-sheet theories are. Thus it is not 
clear what information can been safely extracted from this approach; clarifying this will be the goal of the 
present paper. To do so, we restrict ourself to renormalizable world-sheet interactions which correspond 
to the tachyon and gauge field backgrounds that allows us to use the whole machinery of the renormalization 
group (RG). As to the renormalization conditions, our main requirement which will  
replace their missing will be universality. In other words, we believe in structures in 
BIOST actions which are independent of a renormalization scheme used for world-sheet theories. 

Our conventions and some features of the Witten background independent open string theory that 
are relevant to our discussion are the following:

 (i) In the bosonic case, the world-sheet action is given by
\begin{equation}\label{ac}
S=S_0+S_b
\quad,
\end{equation}
where the bulk action $S_0$ is \footnote{For the sake of simplicity, we use the matrix notations 
here and below.}
\begin{equation}\label{bac}
S_0=\frac{1}{4\pi}\int_D\,d^2z
\left(\pd X^{\text{\tiny T}}\bar\pd X
+4\,b\bar\pd c\,\,+\,\,\text{c.c.}\,\right)
\quad,
\end{equation}
while the boundary action $S_b$ is 
\begin{equation}\label{boac}
S_b=\oint_{\pd D}d\theta\,V=a+\frac{1}{8\pi}\oint_{\pd D}d\theta\left(X^{\text{\tiny T}}t X
+i\dot X^{\text{\tiny T}}f X \right)
\quad.
\end{equation}
Our notations are usual: $D$ is the disk (string world-sheet) whose coordinates are $(z,\bar z)$; $\pd D$ is 
its boundary parameterized by $\theta$; $X^i$'s are the matter fields that 
map the world-sheet into space-time with metric $\delta_{ij}$; $b$ and $c$ are the anti-ghost 
and ghost fields; $a,\,t_{ij},f_{ij}$ are some parameters (coupling constants). 

In fact, the action as it is written in above is the ordinary bosonic open string 
$\sigma$-model action with special profiles for the background tachyon and gauge fields.

 (ii) The partition function of the mater sector 
\begin{equation}\label{pf}
Z(a,t,f)=\int{\cal D}X\,\exp{\left(-S\right)}
\end{equation}
is simply found through the partial differential equations (see \cite{W,LW} for details). In our case, 
the result of formal manipulations can be written as  
\begin{equation}\label{pde}
\frac{\pd \ln Z}{\pd t}=-\frac{1}{4}\langle X^{\text{\tiny T}}(\theta)X(\theta)\rangle
\quad,\quad
\frac{\pd \ln Z}{\pd f}=-\frac{i}{4}\langle \dot X^{\text{\tiny T}}(\theta)X(\theta)\rangle
\quad,
\end{equation}
with the exact Green's function $\langle X(z)X (z')\rangle$ evaluated at boundary points \cite{ LW}
\begin{equation}\label{pr}
\langle X^{\text{\tiny T}}(\theta)X (\theta')\rangle=
\frac{2}{t}+
2\sum_{k=1}^{\infty}
\frac{\ep^{ik(\theta-\theta')}}{k+t+kf}\,+\,
\frac{\ep^{-ik(\theta-\theta')}}{k+t-kf}\
\quad.
\end{equation}
The equations \eqref{pde} can be solved by
\begin{equation}\label{Z}
Z(a,t,f)=\left(\det\, t\right)^{-\oh}\,\ep^{-a}\,\prod_{k=1}^{\infty}\det\left(1+f+k^{-1}t\right)^{-1}
\quad.
\end{equation}
Before continuing our discussion, let us make a couple of general comments: (1) by the differential equations 
$Z$ is defined up to an overall normalization constant $Z(0)$ which we omit. Moreover, $Z(0)$ may 
be in general divergent. (2) The expression \eqref{Z} is purely formal as the infinite product is divergent. 
We postpone discussion of this complication until section 2.

(iii) The background independent open string theory action proposed by Witten based on 
the Batalin-Vilkovisky formalism is defined as \cite{W}
\begin{equation}\label{W}
dS_w=\oh\oint_{\pd D}\oint_{\pd D}d\theta d\theta'\,\langle \,dO(\theta)\,\{Q,O\}(\theta')
\,\rangle
\quad,\quad 
\langle\,\dots\,\rangle=\int{\cal D}X{\cal D}c{\cal D}b\,\exp{\left(-S\right)}
\quad,
\end{equation}
where $O(\theta)=cV(\theta)$; $c$ is a tangential component of the ghost field at the boundary. 
$Q$ means the BRST operator coming from the bulk world-sheet action i.e., in other words, from 
closed string and as a consequence it does not depend on the couplings. Eq.\eqref{W} presents  
a set of partial differential equations which after integration results in the boundary string field theory 
action. According to general arguments \cite{W,LW}, the solution is given by \footnote{We do not write down 
$V$ explicitly leaving its discussion to section 3.} 
\begin{equation}\label{solW}
S_w=\left(1+V^i\frac{\pd}{\pd \text{x}^i}\right)Z
\quad,
\end{equation}
where $V^i$ is a vector field on the space of couplings $\text{x}^i$.

At this point, let us mention that there is a very similar relation between $S$ and $Z$ \cite{S1} namely,
\begin{equation}\label{S}
S_s=\left(1+\beta^i\frac{\pd}{\pd \text{x}^i}\right)Z
\quad,
\end{equation}
where $\beta^i$ is the RG $\beta$-function. It was checked within the perturbation theory up to 
the second order and then it was suggested that it is exact in all orders. In fact, \eqref{S} assumes a
different definition of the BRST operator because it now depends on the couplings (see \cite{S1} for further 
details). Thus one has to consider this as the second definition of the BIOST action. 
So far, it is unknown which action is better. Since the normalization conditions within 
the world-sheet renormalization group are missing it makes no sense to compare these actions. 
What we can compare is some universal structures and we will see in section 3 that these structures 
turned out to be the same in the both cases, so from our point of view these actions are equivalent.

Finally, let us note that both actions are defined modulo the normalization which can be fixed by 
comparing the on-shell three tachyon amplitude computed from these actions with the amplitude computed 
via the cubic open string field theory (see \cite{GSen}).  

(iv) In the fermionic case we follow the NSR formalism, so the world-sheet action is given by
\begin{equation}\label{sac}
\hat S=\hat S_0+\hat S_b
\quad,
\end{equation}
where the bulk action $\hat S_0$ is 
\begin{equation}\label{sbac}
\hat S_0=\frac{1}{4\pi}\int_D\,d^2z
\left(\pd X^{\text{\tiny T}}\bar\pd X+\Psi^{\text{\tiny T}}\bar\pd\Psi
+4\,b\bar\pd c\,\,+4\,\beta\bar\pd\gamma\,\,+
\,\,\text{c.c.}\,\right)
\quad,
\end{equation}
while the boundary action $\hat S_b$ is \footnote{We got such a form by using one-dimensional SUSY and 
the corresponding bosonic action. One can also start from the Witten vertex operator \cite{susyT} 
and then integrating away the auxiliary fields get the same result for a linear tachyon 
profile $T(X)=uX$ (see \cite{susy} for details). Note that it gives $t=u^2$.} 
\begin{equation}\label{sboac}
\hat S_b=\oint_{\pd D}d\theta\,\hat V=a
+\frac{1}{8\pi}\oint_{\pd D}d\theta
\left(X^{\text{\tiny T}}t X+\psi^{\text{\tiny T}}t\pd^{-1}_{\theta}\psi
+i\dot X^{\text{\tiny T}}f X-i\psi^{\text{\tiny T}}f\psi \right)
\quad,
\end{equation}
with $\Psi\vert_{\pd D}=\bar\Psi\vert_{\pd D}=\psi$.

The boundary action \eqref{sboac} is a natural supersymmetrization of the bosonic boundary action 
\eqref{boac}. The supersymmetry transformation acts on the fields in the standard way
\begin{equation}\label{susy}
\delta X=\psi\varepsilon 
\quad,\quad
\delta\psi=-\dot X\varepsilon
\quad.
\end{equation}

As in the bosonic case the partition function of the matter sector 
\begin{equation}\label{spf}
\hat Z(a,t,f)=\int{\cal D}X{\cal D}\Psi{\cal D}\bar\Psi\,\exp{\left(-\hat S\right)}
\end{equation}
can be also found through the partial differential equations. This time they are given by 
\begin{align}
\frac{\pd \ln \hat Z}{\pd t}&=-\frac{1}{4}\langle X^{\text{\tiny T}}(\theta)X(\theta)\rangle
-\frac{1}{4}\langle \psi^{\text{\tiny T}}(\theta)\pd^{-1}_{\theta}\psi(\theta)\rangle
\quad,\quad \label{pde1}\\
\frac{\pd \ln \hat Z}{\pd f}&=-\frac{i}{4}\langle \dot X^{\text{\tiny T}}(\theta)X(\theta)\rangle+
\frac{i}{4}\langle \psi^{\text{\tiny T}}(\theta)\psi(\theta)\rangle
\quad,\label{pde2}
\end{align}
with $\langle X^{\text{\tiny T}}X\rangle$ and $\langle \psi^{\text{\tiny T}}\psi\rangle$ being the exact 
Green's functions evaluated at boundary points. The first has been already given by \eqref{pr}. As to the 
second, it is simply 
\begin{equation}\label{fpr}
\langle \psi^{\text{\tiny T}}(\theta)\psi (\theta')\rangle=
2i\sum_{r=\oh}^{\infty}\frac{r\,\ep^{ir(\theta-\theta')}}{r+t+rf}\,-\,
\frac{r\,\ep^{-ir(\theta-\theta')}}{r+t-rf}\
\quad.
\end{equation}
Since we consider the Neveu-Schwarz sector $r$ is a half integer number.

Formally, the equations \eqref{pde1}-\eqref{pde2} can be solved by
\begin{equation}\label{sZ}
\hat Z(a,t,f)=\left(\det\, t\right)^{-\oh}\,\ep^{-a}\,
\prod_{k=1}^{\infty}\det\left(1+f+k^{-1}t\right)^{-1}\,
\prod_{r=\oh}^{\infty}\det\left(1+f+r^{-1}t\right)
\quad.
\end{equation}
Once more, before continuing our discussion, let us pause to make a couple of general comments: (1) by the 
differential equations $\hat Z$ is defined up to an overall additive normalization constant $\hat Z(0)$ 
we omit. (2) The expression \eqref{sZ} is purely formal as the infinite products are divergent. In 
general, it is natural to expects that SUSY may help us with fixing the problem i.e., the divergences 
cancel each other. The point however is whether this eliminates all divergencies or only some of them. 
We do know one thing about one-dimensional SUSY \cite{AT}. Whatever it does, it can not completely 
eliminate logarithmic divergences. In section 2, we will see whether it happens in the problem at hand. 

(v) Unfortunately, in the fermionic case the notion of the BIOST action as it is given by 
\eqref{W} is still missing. Following \cite{S1}, it seems reasonable to try 
\begin{equation}\label{sS}
\hat S_s=\left(1+\beta^i\frac{\pd}{\pd \text{x}^i}\right)\hat Z
\quad.
\end{equation}
We need to make a couple of remarks here. First, every time the actions are defined as  functions of
the renormalized couplings. Second,  a proposal \eqref{sS} differs from another recent proposal of 
\cite{susy} namely, 
\begin{equation}\label{kut}
\hat S=\hat Z 
\quad.
\end{equation}
This definition inspired by \cite{AT} coincides with \eqref{sS} only at the RG fixed points. A natural 
question to pose is whether these definitions lead to the same result. Again, since the renormalization 
conditions are missing it makes not a lot of sense to compare them. What we can compare is some universal 
structures. We postpone discussion of this issue until section 3.

The outline of the paper is as follows. We start in section 2 by evaluating the partition functions in different 
renormalization schemes to look for universal structures in them. We did so for the tachyon and gauge 
field backgrounds in the both, bosonic and fermionic, cases. Thus, this can be also  considered as 
a generalization of \cite{GS,KMM,susy} to the case of non-zero gauge field \footnote{A recent discussion 
that has overlap with what we describe in the case of bosonic string is due to \cite{Ok}. It 
follows the original paper \cite{W}, however, without discussing ambiguity of the results. Note that 
earlier discussion is due to \cite{LW}.}. Section 3 will present the 
application of our results 
to the BIOST actions. We show that from our point of view (universality) there is no difference between 
two definitions of the action in the bosonic case. Both of them result in the same expression which 
coincides with the result of \cite{GS,KMM} for vanishing gauge field. We also probe our 
proposal \eqref{sS} for superstring and show that it gives the same potential as in \cite{susy,susyZ}. Section 
4 will present our conclusions and remarks.

%\vspace{-5cm}
%_______________________     Section 2 _________________
%\vspace{-.75cm}
\section{ Evaluation of the Partition Function} 
\renewcommand{\theequation}{2.\arabic{equation}}
\setcounter{equation}{0}
 
\subsection{The bosonic string}

We now turn to the problem of the evaluation of the partition function. Actually, the  
 expression for $Z$ as we've already emphasized is purely formal since it is divergent due to the infinite 
product. To make sense of the partition function, one has to introduce a regularization procedure. In papers
\cite{W, LW}, it was done by introducing the normal ordering prescription or short-distance cut-off. 
In general, dealing with functional determinants means that it is more advantageous to use 
the so-called $\zeta$-function regularization which is based on the Riemann 
$\zeta$-function \footnote{ The $\zeta$-function is defined by 
$\zeta (z,q)=\sum_{n=0}^{\infty}\frac{1}{(n+q)^z}$, with $\zeta(z)=\zeta(z,1)$.}. Note that 
this was the case in \cite{FT} where the partition function without the tachyon field was found. So, let us begin 
with the $\zeta$-function regularization scheme.

The expression \eqref{Z} can be rewritten as 
\begin{equation}\label{Z1}
Z(a,t,f)=\left(\det\, t\right)^{-\oh}\,\exp\left(-a-\zeta (0)\tr\ln (1+f)-
\sum_{k=1}^{\infty}\tr\ln\left(1+k^{-1}\nv\right)\right)
\quad.
\end{equation}
Expanding $\ln\left(1+k^{-1}\nv\right)$ in a power series and collecting terms, one finds
\begin{equation}\label{Z2}
Z(a,t,f)=\left(\det\, t\right)^{-\oh}\,\exp\left(-a-\zeta (0)\tr\ln (1+f)-\zeta (1)\tr\left(\nv\right)+
\sum_{n=2}^{\infty}\frac{(-)^n}{n}\zeta (n)\tr\left(\nv\right)^n\right)
\,\,.
\end{equation}
Using also $\zeta (0)=-\oh$, this becomes
\begin{equation}\label{Z3}
Z(a,t,f)=\left(\det \, u\right)^{-\oh}\,\exp\left(-a-\zeta (1)\tr\,u+
\sum_{n=2}^{\infty}\frac{(-)^n}{n}\zeta (n)\tr\,u^n\right)
\quad,
\end{equation}
with 
\begin{equation}\label{u}
u=\nv
\quad.
\end{equation}
If we closely look at the result \eqref{Z3}, it is clear that $Z$ is divergent as 
anticipated \footnote{Note that $\zeta (1)=\infty$. Moreover, $z=1$ is the only singular 
point of $\zeta (z)$.} and, moreover, the dependence on $(t,f)$ enters 
only  through a special combination $\nv$. We renormalize the couplings via the {\it minimal} subtraction 
\begin{equation}\label{RG}
a=a'-\zeta(1)\tr\,u'
\quad,\quad
u=u'
\quad,\quad
\text{with}
\quad\quad
t=t'
\quad,\quad
f=f'
\quad,
\end{equation}
where $(a,t,f)$ are the bare couplings and $(a',t',f')$ are the renormalized ones. 

Next we use the 
renormalization \eqref{RG}, and write the renormalized partition function, $Z'(a', t',f')\equiv Z(a,t,f)$, in 
terms of the new variables $(a',u')$
\begin{equation}\label{Z4}
Z'(a',u')=\left(\det \,u'\right)^{-\oh}\,\exp\left(-a'+
\sum_{n=2}^{\infty}\frac{(-)^n}{n}\zeta (n)\tr\left(u'\right)^n\right)
\quad.
\end{equation}

To simplify further, we will use the following matrix identity 
\begin{equation}\label{GR}
\ln\Gamma (1+M)=\sum_{n=2}^{\infty}\frac{(-)^n}{n}\zeta (n)\,M^n-\gamma\,M
\quad,
\end{equation}
where $M$ is a matrix; $\gamma$ denotes the Euler's constant. Note that it is simply a matrix 
generalization of the identity for the logarithm of the $\gamma$-function \cite{GR}. Thus we end up with 
the renormalized partition function  

\begin{equation}\label{Z5}
Z'(a',u')=\left(\det \,u'\right)^{-\oh}\,\exp\left(-a'+\gamma\,\tr\,u'\right)\det\Gamma(1+u')
\quad.
\end{equation}
Note that we have also used $\det M=\exp\left(\tr\ln M\right)$ to get \eqref{Z5}. 

%___________________________________________________________________________
To complete the story, let us look more closely at what happens within other regularizations. The 
regularization that appeared in \cite{LW} looks a little bit cumbersome for the problem of interest. 
Moreover, it assumes that the boundary propagator \eqref{pr} is modified to 
\begin{equation}\label{pr1}
\langle X^{\text{\tiny T}}(\theta)X (\theta')\rangle=
\frac{2}{t}+
2\sum_{k=1}^{\infty}
\frac{\ep^{ik(\theta-\theta')}}{k+\left(t+kf\right)\ep^{-\epsilon k}}\,+\,
\frac{\ep^{-ik(\theta-\theta')}}{k+\left(t-kf\right)\ep^{-\epsilon k}}\
\quad,
\end{equation}
where $\epsilon$ is the short distance cut-off. It is clear that at vanishing $t$ and $f$ the 
propagator \eqref{pr1} becomes divergent. It poses a natural question whether such a divergence 
is relevant or not \footnote{In fact, the regularization prescription of Li-Witten is more involved. It includes 
nonlocal boundary interactions that do not lead to short distance divergencies, so one does not need the 
regularized propagator for vanishing $t$ and $f$. However at some point it assumes that the interactions 
become local that can run into difficulty.}. 
Instead of answering this question, let us propose another regularization which 
avoids the latter problem. Moreover, it seems a little bit simpler for explicit calculations. In this 
regularization scheme \footnote{In recognition of this, we henceforce denote the parameters 
as $({\bf a},\bf{t},{\bf f})$. For zero value of the tachyon field, this regularization was 
proposed in \cite{T,AT}.}, the propagator is given by 
\begin{equation}\label{pr2}
\langle X^{\text{\tiny T}}(\theta)X (\theta')\rangle=
\frac{2}{t}+
2\sum_{k=1}^{\infty}\ep^{-\epsilon k}\left(
\frac{\ep^{ik(\theta-\theta')}}{k+{\bf t}+k{\bf f}}\,+\,
\frac{\ep^{-ik(\theta-\theta')}}{k+{\bf t}-k{\bf f}}\right)
\quad,
\end{equation}
Now the equations \eqref{pde} can be solved by
\begin{equation}\label{Z6}
Z({\bf a} ,{\bf t} ,{\bf f},\epsilon)=\left(\det {\bf t}\right)^{-\oh}\,\ep^{-{\bf a}}\,
\exp\left(-\sum_{k=1}^{\infty}\ep^{-\epsilon k}\tr\,\ln\left(1+k^{-1}{\bf t}+{\bf f}\right)\right)
\quad.
\end{equation}

The preceding calculations can be generalized without difficulty to this case \footnote{Some useful formulae 
can be found in appendix of \cite{AT}.}. The partition function takes the form 
\begin{equation}\label{Z7}
Z({\bf a} ,{\bf t} ,{\bf f},\epsilon)=
\left(\det {\bf u}\right)^{-\oh}\,
\exp\left(-{\bf a}
+\ln\epsilon\,\tr\,{\bf u}
-\frac{1}{\epsilon}\tr\ln\left(1+{\bf f}\right)
+
\sum_{n=2}^{\infty}\frac{(-)^n}{n}\zeta(n)\tr\,{\bf u}^n
+O(\epsilon)
\right)
\quad,
\end{equation}
in which ${\bf u}=\frac{1}{1+{\bf f}}{\bf t}$. Before we start a comparison of the above result and the result we 
obtained in the framework of the $\zeta$-function regularization, let us make a simple remark. 
In leading order, $\zeta (1)$ may be identified with $-\ln\epsilon$ according to 
$\sum_{n=1}^{\infty}\frac{1}{n}\,\ep^{-\epsilon n}=-\ln\epsilon +O(\epsilon)$. Thus the main novelty 
which is now visible in  Eq.\eqref{Z7} is the appearance of the  power divergence. We again use 
the {\it minimal} subtraction to renormalize the partition function \footnote{See also \cite{T}.}
\begin{equation}\label{RG2}
{\bf a}={\bf a}'+\ln\epsilon\,\tr\,{\bf u}'-
\frac{1}{\epsilon}\tr\ln\left(1+{\bf f}'\right)
\quad,\quad
{\bf u}={\bf u}'
\quad,\quad
\text{with}
\quad
{\bf t}={\bf t}'
\quad,\quad
{\bf f}={\bf f}'
\quad.
\end{equation}
As a result, we have 
\begin{equation}\label{Z8}
Z'({\bf a}',{\bf u}')=\left(\det \,{\bf u}'\right)^{-\oh}\,
\exp\left(-{\bf a}'+\gamma\,\tr\,{\bf u}'\right)\det\Gamma(1+{\bf u}')
\quad.
\end{equation}

In fact, the renormalization as it is done in above is incomplete. The missing point is the renormalization 
conditions which fix the finite part of $Z'$ removing ambiguities due to a particular renormalization scheme. 
Unfortunately, it is not known what these conditions are. Unfortunately, we don't have something new 
to say here. Nevertheless, a natural question which comes to mind to address is whether 
something universal can be found within at least two 
regularization schemes we use. It turns out that this question has a natural answer. First, let us note that 
under a scale transformation $\epsilon\rightarrow\lambda\epsilon$, the expressions \eqref{Z3} and 
\eqref{Z7} transform as \footnote{Here we assume that $\zeta(1)=-\ln\epsilon +O(\epsilon)$.}
 \begin{equation}\label{Z9}
Z(\text{a,t,f})\rightarrow\ep^{\ln\lambda\,\tr\text{u}}Z(\text{a,t,f})
\quad,
\end{equation}
where $\text{a}=(a,{\bf a}),\,\text{t}=(t,{\bf t}),\,\text{f}=(f,{\bf f})$, here and below. By using 
the {\it minimal} subtraction, the renormalized partition function takes the form
\begin{equation}\label{Z10}
Z'(\text{a}',\text{t}',\text{f}')=
\left(\det \,\text{u}'\right)^{-\oh}\,
\exp\left(-\text{ a}'+c\,\tr\,\text{u}'\right)
\det\Gamma(1+\text{ u}')
\quad,
\end{equation}
with an arbitrary constant $c$. This implies that the partition function is a scheme-dependent object. 
Note that it makes no sense to set $c$ equal to a certain number as far as the normalization conditions 
are missing.

%__________________________        B E T A      _______________________
Before continuing our discussion, we will make a detour and rederive some basic results on the RG 
$\beta$-functions in the problem of interest. To do so, let us introduce an arbitrary quantity $\mu$ 
with the dimension of a mass  with respect to the world-sheet theory and some ultraviolet cut-off 
$\Lambda$. Then, Eq.\eqref{RG2} can be  rewritten as 
\begin{equation}\label{RG3}
\frac{{\bf a}}{\mu}
={\bf a}'+\ln\frac{\mu}{\Lambda}\,\,\tr\,{\bf u}'-
\frac{\Lambda}{\mu}\,\,\tr\ln\left(1+{\bf f}'\right)
\quad,\quad
\frac{{\bf t}}{\mu}={\bf t}'
\quad,\quad
{\bf f}={\bf f}'
\quad,
\end{equation}
in which $\epsilon=\frac{\mu}{\Lambda}$ and all renormalized couplings are explicitly dimensionless 
with respect to the world-sheet theory. 

It is well-known that the scale transformations $\mu\rightarrow\mu (1+\sigma)$ with an infinitesimal 
parameter $\sigma$ result in the corresponding shifts of the renormalized couplings i.e., 
$\text{x}^i\rightarrow\text{x}^i+\sigma\beta^i$. The function $\beta^i$ is called the RG 
$\beta$-function. A simple algebra shows that in the problem at hand
\begin{equation}\label{beta2}
\beta_{\bf a}=-\tr\,{\bf u}'-{\bf a}'
\quad,\quad
\beta_{\bf t}=-{\bf t}'
\quad,\quad
\beta_{\bf f}=0
\quad.
\end{equation}
It is a straightforward matter of working out the calculations within the $\zeta$-function regularization. 
The result is the same as in above. 

Now consider RG fixed points which are defined by  $\beta_{\text {x}}=0$ \footnote{We do not 
consider $\beta_{\text {x}}=\infty$.}. It is easy to see at least one family of the solutions  
\begin{equation}\label{sol}
\text{a}'=0
\quad,\quad
\text{t}'=0
\quad,\quad 
\text{f}'=\text{fixed}
\end{equation}
that is parameterized by the antisymmetric tensor $\text{f}'$. 

It is also easy to see that at the vicinity of these solutions the partition function shows the critical behavior. 
Indeed, for example setting t' to be diagonal $\text{t}'=\text{diag}(t_1,\dots ,t_n)$, one finds
\begin{equation}\label{Z11}
Z'(\text{a}',\text{t}',\text{f}')\sim \prod_{i=1}^n t_i^{-\oh} 
\quad,\quad
\text{as}
\quad
t_i\rightarrow 0
\quad.
\end{equation}
 
Since it is usual to say that the leading asymptotic of the partition function at the vicinity of the 
critical points should be universal we have at least one explicit asymptotics in the problem of interest
\begin{equation}\label{Z12}
Z'(\text{a}',\text{t}',\text{f}')=\left(\det \,\text{u}'\right)^{-\oh}\,\ep^{-\text{ a}'}
\quad
\text{for}\quad \text{t}'\sim 0
\end{equation}
At this point we should stress that we relax the requirement $\text{a}'\sim 0$ and include the factor 
$\ep^{-\text{a}'}$ in $Z'$. Motivations for doing so come from our main suggestion which is to replace  
unknown normalization conditions by universality. Indeed, Eq.\eqref{Z10} shows that such a factor is 
universal while the ambiguity due to $c$ drops from the above asymptotics that is in agreement with our 
expectations.
 
Let us conclude this subsection by making some remarks. 

(i) Our result \eqref{Z10} passes through the 
requirement of Ref.\cite{LW} which in the case of interest is
\begin{equation}\label{test}
Z'(\text{a}',\text{t}',\text{f}')\vert_{\text {f}'=0}=Z'(\text {a}',\text{t}')
\quad.
\end{equation} 
 This fact is evident since the renormalized couplings for the tachyon and  gauge field enter only 
through a special combination $\frac{1}{1+\text{f}'}\text{t}'$. So, Eq.\eqref{test} can not be considered as 
complete renormalization conditions. 

(ii) A simple algebra show that the result \eqref{Z10} is in a complete agreement with 
earlier computations \cite{FT, N}. Indeed, setting $\text{t}'=0$ one finds that the 
only $\text{f}'$-dependent contribution comes from the first factor and it is $\sqrt{\det (1+\text{f}')}$. The 
divergence due to $\det \text{t}'$ is related to the fact that in this case the fields $X^\mu$ 
have zero modes and  must be interpreted as the space-time volume (see \cite{LW} for details). As 
a result, $Z'$ provides the Born-Infeld action. 

(iii) There exists another interesting possibility to fulfill the requirement of universality. To see what 
happens, let us introduce a general constant space-time metric $g_{ij}$ for the bulk action \eqref{bac}. 
This results in a minor modification of all story as $\text{u}$ now 
becomes $\text{u}=\frac{1}{g+\text{f}}\text{t}$. For $g=0$, we have $\tr\,\text{u}\equiv 0$ that allows 
us to conclude that $c$ in \eqref{Z10} drops out. So, 
it makes sense to keep all terms in the expression \eqref{Z10}. Using 
\begin{equation}\label{GR2}
\ln\Gamma (1+M)=\oh\ln\left(\frac{\pi M}{\sin\pi M}\right)-
\sum_{n=1}^{\infty}\frac{1}{2n+1}\zeta (2n+1)\,M^{2n+1}-\gamma\,M
\quad,
\end{equation}
that is a matrix generalization of the identity for the logarithm of the $\gamma$-function \cite{GR},
we end up with the following expression for the partition function \footnote{We omit overall numerical 
factors in expressions for the partition functions as they have no meaning.}
\begin{equation}\label{Z14}
Z'(\text{a}',\text{t}',\text{f}')=
\left(\det \,\sin (\pi\frac{1}{\text{f}'}\text{t}')\right)^{-\oh}\,
\exp\left(-\text{ a}'\right)
\quad.
\end{equation}

At this point, it is not difficult to see some analogy with the Seiberg-Witten description of open string 
theory in the presence of a constant $B$-field \cite{SW}.  Indeed, we have for the open string metric and 
parameter of noncommutativity $G^{-1}=0$ and 
$\theta_n= \frac{1}{B}=\frac{1}{\text{f}'}$, respectively. However we do not take the 
$\alpha'\rightarrow 0$ as we wish to hold $\alpha'$-dependence.

%________________________   S U S Y   ________________________________
\subsection{The fermionic string}

We will now extend our analysis to the problem of the evaluation of the partition function, including the 
integration over fermions. As we've already emphasized the expression \eqref{sZ} is purely formal since 
it may be  divergent. To make sense of the partition function, one has to introduce a regularization 
procedure. Following subsection 2.1, we begin with the $\zeta$-function regularization scheme. 

First, we rewrite \eqref{sZ} as 
\begin{equation}\label{sZ1}
\begin{split}
\hat Z(a,t,f)&=\left(\det\, t\right)^{-\oh}\,
\exp\left(-a+
\left[\zeta(0,\oh )-\zeta (0)\right]\tr\ln (1+f)\right) \\
&\phantom{=}\times
\exp\left(-\sum_{k=1}^{\infty}\tr\ln\left(1+k^{-1}\nv\right)
+\sum_{r=\oh }^{\infty}\tr\ln\left(1+r^{-1}\nv\right)
\right)
\quad.
\end{split}
\end{equation}
Expanding the logarithms in power series and collecting terms, one finds
\begin{equation}\label{sZ2}
\begin{split}
\hat Z(a,t,f)&=\left(\det\, t\right)^{-\oh}\,
\exp\left(-a+
\left[\zeta(0,\oh )-\zeta (0)\right]\tr\ln (1+f)
+\left[\zeta(1,\oh )-\zeta (1)\right]\tr\nv
\right) \\
&\phantom{=}\times
\exp\left(
\sum_{n=2}^{\infty}\frac{(-)^n}{n}
\left[\zeta (n)-\zeta(n,\oh )\right]\tr\left(\nv\right)^n
\right)
\quad.
\end{split}
\end{equation}
Finally, using $\zeta (n,\oh )=(2^n-1)\zeta(n)$ and introducing $u$, we get
\begin{equation}\label{sZ3}
\hat Z(a,u)=
\left(\det \, u\right)^{-\oh}\,\exp\left(-a+
\sum_{n=2}^{\infty}\frac{(-)^n}{n}(2-2^n)\zeta (n)\tr\,u^n
\right)
\quad. 
\end{equation}
The main novelty now is that the partition function became finite. So, we do not need to subtract the 
logarithmic divergence anymore. We postpone the introduction of the renormalized couplings for a 
moment. At this point a comment is in order.

Using \eqref{Z4}, a small computation shows that the fermionic and bosonic 
partition functions obey 
\begin{equation}\label{id}
  \hat Z(a,u)=\frac{\left(Z'(a,u)\right)^2}{Z'(a,2u)}
\quad.
\end{equation} 
This is different from what is recently found in \cite{susy}. The difference is due to a factor $4^{\tr u}$. 
Due to the appearance of $\tr u$ it immediately comes to mind to make a double check by using another 
regularization scheme as the term with $\tr u$ was the origin of all our troubles in subsection 2.1. 

Let us compute the partition function within a regularization scheme which contains an explicit short 
distance cut-off $\epsilon$. For our purposes it turns out enough to define the regularized propagator 
for the $X$'s as \eqref{pr2} while for the fermions as 
\begin{equation}\label{pr3}
\langle \psi^{\text{\tiny T}}(\theta)\psi (\theta')\rangle=
2i\sum_{r=\oh}^{\infty}
\ep^{-\epsilon r}\left(
\frac{r\,\ep^{ir(\theta-\theta')}}{r+{\bf t}+r{\bf f}}\,-\,
\frac{r\,\ep^{-ir(\theta-\theta')}}{r+{\bf t}-r{\bf f}}\right)
\quad.
\end{equation}

Now the equations \eqref{pde1}-\eqref{pde2} can be solved by
\begin{equation}\label{sZ6}
\hat Z({\bf a} ,{\bf t} ,{\bf f},\epsilon)=\left(\det {\bf t}\right)^{-\oh}\,\ep^{-{\bf a}}\,
\exp\left(-\sum_{k=1}^{\infty}\ep^{-\epsilon k}\tr\,\ln\left(1+k^{-1}{\bf t}+{\bf f}\right)
+\sum_{r=\oh }^{\infty}\ep^{-\epsilon r}\tr\,\ln\left(1+r^{-1}{\bf t}+{\bf f}\right)
\right)
\quad.
\end{equation}

The preceding calculations can be generalized without difficulty to this regularization scheme. So we have 
\begin{equation}\label{sZ7}
\hat Z({\bf a},{\bf u},\epsilon)=
\left(\det \, {\bf u}\right)^{-\oh}\,\exp\left(-{\bf a}+\ln 4\,\tr\,{\bf u}+
\sum_{n=2}^{\infty}\frac{(-)^n}{n}(2-2^n)\zeta (n)\tr\,{\bf u}^n
\right)
\quad. 
\end{equation}
Thus the partition function is again finite but it differs from the former one by a factor $4^{\tr\,{\bf u}}$. 
On the one hand, it sounds good as it is in harmony with the result of \cite{susy}. On the other hand, it 
poses a natural question about universality we have already discussed in subsection 2.1 as a general form 
of $\hat Z$ turns out to be 
\begin{equation}\label{sZ8}
\hat Z(\text {a},\text{u})=
\left(\det \, \text {u}\right)^{-\oh}\,\exp\left(-\text{ a}+\hat c\,\tr\text{u}+
\sum_{n=2}^{\infty}\frac{(-)^n}{n}(2-2^n)\zeta (n)\tr\,\text{u}^n
\right)
\quad,
\end{equation}
where $\hat c$ is an arbitrary constant. 

It seems reasonable to set $\text{a}=0$ as this coupling has no meaning in the fermionic case. Now, 
Eq.\eqref{RG3} is modified to 
\begin{equation}\label{RG4}
\frac{\text{ t}}{\mu}=\text{t}'
\quad,\quad
\text{f}=\text{f}'
\quad,
\end{equation}
that still allows non-trivial $\beta$-function
\begin{equation}\label{beta3}
\beta_{\text{t}}=-\text{t}'
\quad,\quad
\beta_{\text{f}}=0
\quad.
\end{equation}
The RG fixed points are again a family parameterized by antisymmetric tensor (see \eqref{sol}). 

As the asymptotics at the vicinity of the critical points is believed to be universal we get 
\begin{equation}\label{sZ9}
\hat Z(\text{u}')\sim\left(\det \, \text {u}'\right)^{-\oh} 
\quad, \quad
\text{for}\quad
\text{t}'\sim 0
\quad,
\end{equation}
that is in harmony with our expectations as $\hat c$ drops out.

At this point, let us makes some remarks.

(i) As in the bosonic case, our result for the partition function \eqref{sZ8} simply passes through 
the requirement like \eqref{test}. Thus, the use of that as the renormalization conditions fails.

(ii) As we have discussed at the end of the previous subsection, the asymptotics like \eqref{sZ9} or 
expressions like \eqref{sZ8} at $\text{t}=0$ recover the Born-Infeld action. So, there is a complete 
agreement with earlier computations \cite{sT, MRT}.

(iii)  In the case of a general space-time metric $g$, we can take $g=0$. If we do so,  then the origin of our 
troubles, $\tr\,\text{u}$, identically equals zero. With such a good novelty we can hold all terms in \eqref{sZ8}. 
Some evidence that it is a right way comes from our results and from a general consideration within 
the perturbation theory which shows that these terms appear by finite integrals. Thus the partition function 
becomes 
\begin{equation}\label{sZ14}
\hat Z(\text{a}',\text{t}',\text{f}')=
\left(\det \,\tan(\pi\frac{1}{\text{f}'}\text{t}')\right)^{-\oh}
\quad.
\end{equation}
 To get this form we used the matrix relations: $\sin 2M=2\sin M\cos M$ and $\tan M=\sin M/\cos M$. 

 %\vspace{-5cm}
%_______________________     Section 3 _________________
%\vspace{-.75cm}
\section{ String Field Actions} 
\renewcommand{\theequation}{3.\arabic{equation}}
\setcounter{equation}{0}

Now we come to a key point. What is the background independent string field theory action in 
which we can belive? If the normalization conditions are missing, it makes no sense to rigorously 
speak about $S$. Nevertheless, 
we will try to do so. Our main requirement which replaces the normalization conditions will be 
universality. In other words, we believe in structures which are independent of renormalization schemes
 for the world-sheet theories. 

In  studying the BIOST actions, we begin with the bosonic string. The basic fact we've 
learned in section 2 is that it is a pointless deal to use the expression of $Z'$ computed in a special 
regularization scheme for lack of universality. Instead of this, we can  believe at least in the 
asymptotics \eqref{Z12}. As a result, all linear terms in $\text{t}'$ can be dropped in deriving the 
asymptotics of the background independent string field theory action for $\text{t}'\sim 0$. This leads to 
a significant simplification in the integration of the partial differential equations \eqref{W}. Moreover, as 
we will see in a moment it avoids the problem we mentioned in introduction namely, why $S_w$ is better 
than $S_s$ and vice versa. 

We are now ready to go. The simplest way to get what we need is to consider the $\zeta$-function
regularization scheme. So, taking
\begin{equation}\label{dO}
dO(\theta)=c\left(\frac{1}{2\pi}d\text{a}+\dots\right)(\theta) 
\quad,
\end{equation}
together with 
\begin{equation}\label{Q}
\{Q,O\}(\theta')=c\dot c\left(\frac{1}{2\pi}a+\frac{1}{8\pi}X^{\text{\tiny T}}t X\dots\right)(\theta')
\quad,
\end{equation}
where dots mean terms that are irrelevant, we immediately get the equation
\begin{equation}\label{eq}
\frac{\pd S_w}{\pd a}=\left(-a+t\frac{\pd}{\pd t}\right)Z
\quad.
\end{equation}

In our approximation , it can be solved by 
\begin{equation}\label{W2}
S_{w}=\left(1+a-t\frac{\pd}{\pd t}\right) Z 
\quad.
\end{equation}
Using the expression \eqref{Z3}, we see that $a$ can be  renormalized as \eqref{RG}. Thus the 
background independent string field theory action is a function of the renormalized couplings. At 
this point a couple of remarks is in order.

(i) Note that the right hand side of Eq.\eqref{Q} is due to the ghosts contractions, so it does not depends on the 
details of the matter sector, namely whether the BRST charge depends on the couplings or not.  
From this point of view it is universal. 

(ii) In general, it is a non-trivial consistency check to see that the BIOST action is a 
function on the renormalized couplings. In the framework of another regularization scheme it was 
shown in \cite{LW}. Thus, it appears that this is the case. 

Finally, we have the following asymptotics for the BIOST action 
\begin{equation}\label{W4}
S_w=\left(\det \,\text{u}'\right)^{-\oh}\,\left(1+\text{a}'+\frac{D}{2}\right)\ep^{-\text{ a}'}
\quad\text{for}\quad
\text{t}'\sim 0
\quad.
\end{equation}
Now let us try to see what happens with the action $S_s$ defined in \eqref{S}. Borrowing our 
result for the RG $\beta$-functions \eqref{beta2} from section 2, we immediately get the same expression 
as above. Thus we conclude that the universal structure of $S_w$ and $S_s$ is the same i.e., 
from our point of view there is no difference between these actions. At this point, a natural question to ask 
is what to do with non-universal structures that are responsible for difference. Unfortunately, we do not 
know  what the exact answer is. We can only suggest that these structures may be transformed each into 
another under a set of coupling's redefinitions. 

To compare our result with ``experiment'', we follow the lines of \cite{KMM}. So, we take 
the effective action for the tachyon and gauge field 
\begin{equation}\label{ea}
S_{eff}=\int d^{D}x\sqrt{\det\left(1+F\right)}\,\ep^{-T}\Bigl(1+T+O\left(\pd T,\pd F\right)\Bigr)
\quad,
\end{equation}
where $O(\pd T,\pd F)$ means derivative terms. Next we evaluated it for the profiles whose forms 
are motivated by Eq.\eqref{boac}
\begin{equation}\label{ps}
T(X)=\text{a}'+X^{\text{\tiny T}}\text{t}'X
\quad,\quad
F(X)=\text{f}'
\quad.
\end{equation}
A simple algebra shows that the result is given by \eqref{W4} that is exactly what we need. Note that 
for $F=0$ the same potential was obtained in \cite{GS, KMM}. Moreover, it has recently appeared in 
\cite{Ok}. 

Note that we have not included the derivative terms in \eqref{ea}. The latter is meaningless as this would 
lead to $\text{t}'$ corrections which are over control in the framework of the background independent 
string field theory we consider. Thus it makes no sense to speculate about the tachyon mass. We do not 
control it. What we certainly control is the only potential that is related with the universal structure. 

To conclude our discussion of bosonic string, let us remark that the expression \eqref{Z14} in fact leads to 
the tachyon potential within the noncommutative scalar field theory whose parameter of noncommutativity 
$\theta_n=\frac{1}{\text{f}'}$. A formal proof is straightforward: first we use it to compute $S$; next we 
go along the lines of \cite{SW} \footnote{See also \cite{Cor,Ok} for more details.}. The only difference is 
that in our case $G^{-1}=0$ while $\theta_n$ is arbitrary. The first allows us to suppress all derivative 
terms. As to the second, it allows us to avoid taking a large noncommutativity limit that  shows that the form 
of potential is universal i.e., it is the same for all $\theta_n (\text{f}')$. 

%____________________  susy  S  ____________________________________________
We now turn to the problem of shedding some light on the boundary string field theory action in the 
fermionic case. As we've already mentioned the fermionic counterpart of the Witten's action \eqref{W} is still 
missing, so we suggested to define the action as \eqref{sS}. Our motivation to do so is the following:  as we've 
seen in section 2, there is no difference between two actions $S_w$ and $S_s$ from our point of view i.e., 
universality. So, if the definition of $\hat S_w$  in the framework of the Batalin-Vilkovisky 
formalism is absent, then we can use our definition of $\hat S_s$. The latter is based on a natural 
supersymmetrization of $S_s$ . 

From \eqref{beta3} and \eqref{sZ9}, we get
\begin{equation}\label{sS2}
\hat S_s\sim\left(\det \, \text {u}'\right)^{-\oh} \,\left(1+\frac{D}{2}\right)
\quad, \quad
\text{for}\quad
\text{t}'\sim 0
\quad,
\end{equation}
that formally coincides with the bosonic action \eqref{W4} evaluated at $\text{a}'=0$. 

The comparison of the above result with ``experiment'' is straightforward. Taking the effective action 
\begin{equation}\label{sea}
\hat S_{eff}=\int d^{D}x\sqrt{\det\left(1+F\right)}\,\ep^{-T^2}
\Bigl(1+T^2+O\left(\pd T,\pd F\right)\Bigr)
\quad,
\end{equation}
and evaluating it for the linear profiles \footnote{A constant $\text{a}'$ has been omitted, because it can 
be absorbed in a shift in the $X$'s (as long as $u\not=0$). This is in a complete agreement with our 
conclusions of section 2 where we dropped $\text{a}'$ based on the RG analysis.} 
\begin{equation}\label{sps}
T(X)=uX 
\quad
\text{with}
\quad
\text{t}'=u^2
\quad;\quad
F(X)=\text{f}'
\quad,
\end{equation}
we immediately get a desired result \eqref{sS2}. 

On the other hand, the definition \eqref{kut} gives simply
\begin{equation}\label{kut1}
\hat S\sim\left(\det \, \text {u}'\right)^{-\oh} 
\quad, \quad
\text{for}\quad
\text{t}'\sim 0
\quad.
\end{equation}
So the effective action is now \footnote{For $F=0$, this action has been proposed based on the exactly 
solvable Schroedinger problem \cite{susyZ}. }
\begin{equation}\label{sea2}
\hat S_{eff}=\int d^{D}x\sqrt{\det\left(1+F\right)}\,\ep^{-T^2}
\Bigl(1+O\left(\pd T,\pd F\right)\Bigr)
\quad.
\end{equation}
We should noted that for $F=0$ the same result has been recently found in \cite{susy}. 

At first glance, the expressions \eqref{sea} and \eqref{sea2} are not equivalent. However, because we omit 
the derivative terms we can transform the actions each into another by a field redefinition. Moreover, 
the potentials have a similar profile with the same extrema at $T_{ext}=\{0,\infty\}$. So, from this 
point of view we can not say that the expressions are different. Our conclusion is that the factor 
$\sqrt{\det\left(1+F\right)}\ep^{-T^2}$ in the effective actions is universal. Unfortunately, we have no 
arguments in our disposal to say why $\hat S_s$ is better than $\hat S$ or vice versa.

As in the bosonic case, we do not  include the derivative terms in \eqref{sea},\eqref{sea2} because they 
lead to $\text{t}'$ corrections which are over control. Thus it makes no sense to speculate about the 
tachyon mass. We only control the potential that is related with the universal structure. 

To conclude our discussion of fermionic string, let us discuss the expression \eqref{sZ14} more carefully than 
we have done that for its counterpart in the case of bosonic string where the expression \eqref{Z14} leaded 
us to the potential of the noncommutative scalar field theory. Indeed, our discussion was a little bit cavalry. 
The point is that noncommutative field theories appear from open string in the presence of a constant 
$B$-field only within a very special regularization scheme, the so-called point splitting 
regularization \cite{SW} while we got the noncommutative scalar field theory for all our schemes. Moreover, 
this was also the case in \cite{Cor, Ok}. Examples with the tachyon vertex operator are in some sense 
confusing as any regularization scheme gives the $\ast$-product structure \footnote{We refer to \cite{SW} 
for a review of noncommutative geometry within string theory.}. To be more precise, let us 
recall two famous facts: first, the bosonic tachyon vertex operator is simply $\ep^{ikX}$; second, the 
propagator of $X$'s  evaluated at the boundary of the upper half plane is given by 
\begin{equation}\label{xp}
\langle X^{\text{\tiny T}}(\tau)X(\tau')\rangle=-2\,G^{-1}\ln\left(\tau-\tau'\right)^2\,
+2\pi i\,\theta_n\,\varepsilon(\tau-\tau')
\quad.
\end{equation}
In our notations, $G=1-f^2$, $\theta_n=f/1-f^2$ (see \eqref{pr}). Thus, every time $\theta_n$ appears 
in a such way that  the $\ast$-product structure is correct. However, we run into difficulty as soon as 
we would like to consider other vertex operators, for example, the gauge field vertex operator. The known 
remedy is the use of the point splitting regularization. Now let us see what happens in the fermionic case. 
This time the tachyon vertex operator \footnote{We mean the operator that follows from the 
boundary action \eqref{sboac}.} gets a contribution like $k\psi\pd_\tau^{-1}k\psi\ep^{ikX}$ by the 
fermions $\psi$'s whose propagator evaluated at the boundary of the upper half plane is 
\begin{equation}\label{fp}
\langle \psi^{\text{\tiny T}}(\tau)\psi(\tau')\rangle=4\,G^{-1}\frac{1}{\tau-\tau'}\,
-4\pi i\,\theta_n\,\delta(\tau-\tau')
\quad.
\end{equation}
It is clear that the last term in \eqref{fp} clashes with noncommutativity since $\theta_n$ can
appear not only due to the $\ast$-product. The remedy is again the use of the point splitting 
regularization that automatically fixes the problem. After this is understood, it becomes clear that 
it makes no sense to look for noncommutative field theory to compare its action with the action coming 
from the use of the expression \eqref{sZ14}. First, we have to get rid of the fermionic contribution. 
It is clear that this brings the result to the bosonic form \eqref{Z14}. So, for example, using the 
definition \eqref{kut}, we get 
\begin{equation}\label{nsea}
\hat S_{eff}=\int d^{D}x\,\ast \ep^{-T^2}
\quad.
\end{equation}
We do not include derivative terms as $G^{-1}=0$. 

%\vspace{-5cm}
%_______________________     Section 4 _________________
%\vspace{-.75cm}
\section{ Conclusions and Remarks} 
\renewcommand{\theequation}{4.\arabic{equation}}
\setcounter{equation}{0}
First let us say a few words about the results.

In this work we have discussed what information can be safety extracted from background independent 
off-shell string theory in the case of missing renormalization conditions for the underlying world-sheet 
theories. We propose a principle of universality to somehow compensate this drawback. It states that 
one believes in structures in BIOST actions which are independent of a renormalization scheme used 
for the world-sheet theories. To do so, we considered the tachyon and gauge fields backgrounds and 
carried out computations in the different renormalization schemes for both, bosonic and fermionic strings. 
In the latter case, we also proposed the definition of the BIOST action since the rigorous definition based on the 
Batalin-Vilkovisky formalism is still missing in this case. Our main conclusion is that some asymptotics 
which are responsible for the potentials only obey the principle of universality.   

Let us conclude by making a couple of remarks.

(i) All the results for the potentials can be easily obtained in the framework of the so-called sigma model 
approach to string theory \footnote{See the last reference in \cite{sigma} for a review.} equipped with 
the definitions of the actions. Let us briefly illustrate how this goes in the case of 
superstring. Splitting the integration variable $X$ in the constant and the non-constant part as 
$X(\theta)=x+\xi(\theta)$ and dropping all derivative terms like $\pd T,\,\pd F$ from the boundary 
interactions
\begin{equation}\label{bsigma}
\hat S_b=\oint_{\pd D}d\theta\, \left(T^2(X)+\psi\pd T\pd_\theta^{-1}\psi\pd T(X)+i\dot X A(X)+
\frac{i}{2}\psi^{\text{\tiny T}}F(X)\psi\right)
\quad,
\end{equation}
we immediately get for the partition function
\begin{equation}
\hat Z\sim \int d^D x\, \sqrt{\det (1+F)}\,\ep^{-T^2}
\quad,
\end{equation}
where the Born-Infeld factor is due to the integration over $\xi$ \cite{MRT} and $T$ is real. Thus, in the 
case $\hat S=\hat Z$ we recover the result \eqref{sea2}. On the other hand,
\begin{equation}\label{beta4}
\beta_T=-\oh T+O(\pd T,\pd F)
\quad,\quad
\beta_{A}=0+O(\pd T, \pd F)
\quad.
\end{equation}
together with the definition \eqref{sS} exactly result in Eq.\eqref{sea} i.e., what we got in section 3. The 
expressions for the $\beta$-functions follow from our result \eqref{beta3}. For example, substituting 
$\text{t}'=u^2$ into the first equation of \eqref{beta3}, we get $\beta_u=-\oh u$ which for the linear 
tachyon profile  \eqref{sps} provides a desired result.

On the one hand, an apparent advantage of this approach with respect to BIOST is that it works for any 
profile of the tachyon field \footnote{In fact, it was used in \cite{Cor} to derive the potential in the case of 
bosonic string with a large $B$-field.}. On the other hand, BIOST turns out to be much simpler to analyze 
ambiguity related with derivative terms. 

(ii) The main motivation in studying the tachyon potentials is the phenomenon of 
tachyon condensation \cite{tc}. It has been recently realized that BIOST turns out to be very useful 
to get a correct energetics of this phenomenon \cite{KMM, GSen}. We are not going to discuss all related 
issues here as it is out of scope of the present paper but there is one which we would like to mention. 
It is clear from our analysis that the actions as well as partition functions depend on a universal variable 
$\text {u}=\frac{1}{1+\text{f}}\text{t}$ rather than $\text{t}$ and $\text{f}$. If one calls $\text{a}=\infty,\,\,
 \text{t}=0$ as the vacuum configuration \cite{GSen}, then   
$\text{f}=\text{fixed}$ is irrelevant for this configuration.  This is in harmony with the Sen conjecture that 
at the minimum of the tachyon potential all different values of the gauge field strength $F$ describe 
the same physics \cite{Sen2}.

%__________________       Acknowledgments   ________________________
\vspace{.25cm} {\bf Acknowledgments}

\vspace{.25cm} 
We would like to thank A.A. Tseytlin for a collaboration at an initial stage, and also for useful 
comments. We are also grateful to B. Zwiebach for helpful discussions and H. Dorn for reading the 
manuscript. The work is supported in part by DFG under Grant No. DO 447/3-1 and the European 
Community grant INTAS-OPEN-97-1312.

%__________________                      R E F S                    ______________________

%__________________________________________________________

\end{document}